\documentclass[sigconf]{acmart}

\usepackage{amssymb}
\usepackage{booktabs}
\usepackage{color}
\usepackage{hyperref}
\usepackage{algorithm}
\usepackage{amsmath}
\usepackage{amsfonts}
\usepackage{amsthm}
\usepackage{bbm}
\usepackage{balance}
\usepackage{hyperref} 
\definecolor{mylinkcolor}{RGB}{0,0,0}
\hypersetup{colorlinks,allcolors=mylinkcolor,citecolor=mylinkcolor}
\usepackage[capitalize]{cleveref}

\allowdisplaybreaks

\newcommand{\xhdr}[1]{\vspace{1.0mm}\noindent{{\bf #1.}}\hspace{0.5mm}}

\begin{document}

\title{Scaling choice models of relational social data}

\author{Jan Overgoor}
  \affiliation{\institution{Stanford University}  \city{Stanford} \state{CA}}
  \email{overgoor@stanford.edu}
\author{George Pakapol Supaniratisai}
  \affiliation{\institution{Stanford University}  \city{Stanford} \state{CA}}
  \email{sspkpl@stanford.edu}
\author{Johan Ugander}
  \affiliation{\institution{Stanford University}  \city{Stanford} \state{CA}}
  \email{jugander@stanford.edu}

\begin{abstract}
Many prediction problems on social networks, from recommendations to anomaly detection, 
can be approached by modeling network data as a sequence of relational events and 
then leveraging the resulting model for prediction.
Conditional logit models of discrete choice are a natural approach to modeling relational 
events as ``choices'' in a framework that envelops and extends 
many long-studied models of network formation.
The conditional logit model is simplistic, but it is particularly attractive because it allows for efficient consistent likelihood maximization via negative sampling, something that isn't true for mixed logit and many other richer models. The value of negative sampling is particularly pronounced because choice sets in relational data are often enormous.

Given the importance of negative sampling, 
in this work we introduce a model simplification technique 
for mixed logit models that we call ``de-mixing'', 
whereby standard mixture models of network formation---particularly models that mix local and global link formation---are reformulated to operate their modes over disjoint choice sets.
This reformulation reduces mixed logit models to conditional logit models, 
opening the door to negative sampling while also circumventing other standard challenges with maximizing mixture 
model likelihoods. To further improve scalability,
we also study importance sampling for more efficiently selecting negative samples,
finding that it can greatly speed up inference in both standard and de-mixed models. Together, these steps make it possible to much more realistically model network formation in very large graphs.
We illustrate the relative gains of our improvements on synthetic datasets
with known ground truth as well as a large-scale dataset of public 
transactions on the Venmo platform. 
\end{abstract}

\maketitle

\section{Introduction}

Many modern challenges in mining social data can be cast as modeling the likelihood 
of edges or events between nodes.
Link prediction, the problem of who to recommend a user to friend or connect with~\cite{liben07}, is well-addressed by statistical approaches~\cite{ghasemian19}. Anomaly detection, finding outliers in relational data to identify fraudulent activity~\cite{noble03}, is well-addressed by identifying low-probability events under a statistical model~\cite{akoglu15}.
Recommendation systems examine frequent relationships in a bipartite graph between users and items~\cite{agrawal93}, where model-based approaches again can be highly effective and efficient.
Across these problems there is a common language of \textit{relational events}, which are events involving two or more\footnote{In this work we focus on binary relational events, but recent work extends choice models to hypergraphs to model relations between sets of nodes \cite{benson18,benson18b}} units, viewed as nodes in a graph where events connect these units by edges. Persistent events form static graphs, while transient events (such as in communication, transaction, or consumption data) form dynamic graphs.
Relational event modeling can thus be applied to a large number of data mining applications.

\begin{figure}
    \centering
    \includegraphics[width=0.33\textwidth]{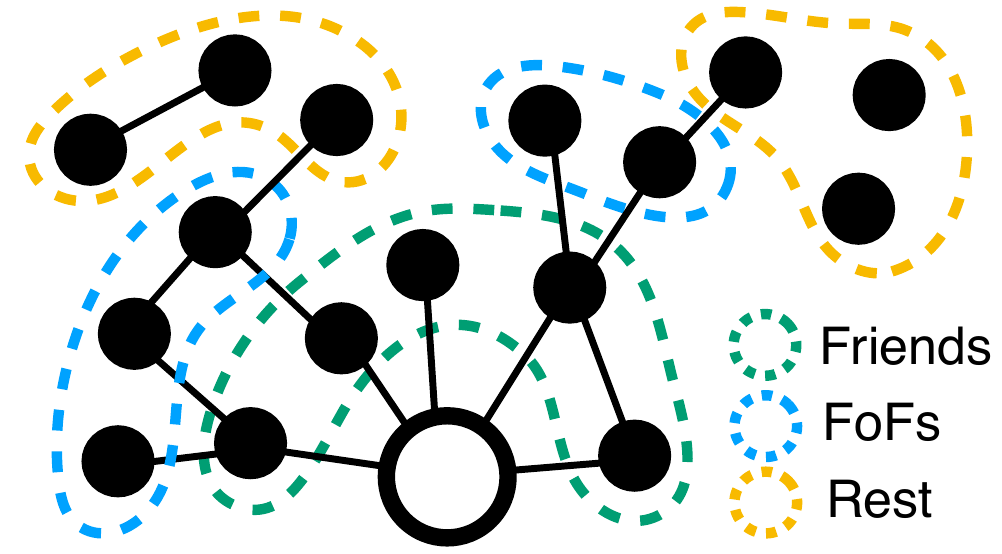}
    \caption{
     Illustration of a choice process in a small network.
     The ego chooses who to interact with, where different choice models may apply to friends, friends-of-friends, and others.
    }
    \label{fig:diagram}
\end{figure}

Discrete choice modeling provides a natural framework for modeling relational events \cite{butts08,stadtfeld17a}.
Each event is viewed as a choice made by one node to involve another node, and modeled based on features of all the alternatives (see Figure~\ref{fig:diagram}).
The conditional logit model of relational events subsumes and extends many existing models of network formation,
such as preferential attachment and triadic closure \cite{overgoor19}.
Yet in network analysis, the ability to work with large datasets is a major concern, 
where modeling relational events as choices raises both practical and conceptual issues.

The first set of issues pertains to the scale of the data.
For a graph containing $n$ nodes, when the originator of the choice is known
  then every \textit{event} represents a choice with $O(n)$ alternatives. 
For large and sparse graphs, the large slates of potential alternatives
makes direct inference intractable.
Existing frameworks for modeling network formation with a logit model, 
  such as SAOMs \cite{snijders10} and REMs \cite{butts08}, are severely restricted in
  the size of the data they can directly handle \cite{lerner19}.
Critically, however, under the conditional logit model the non-chosen alternatives can be sampled 
via a procedure commonly called ``negative sampling'' which produces estimates 
  that are consistent for the estimates on the full data \cite{mcfadden77,train09}.

\xhdr{Availability, mixing, and de-mixing}
Another issue with applying the conditional logit model to large graph datasets is the \textit{availability} assumption 
that the chooser is a rational actor who has complete information about their available options and their features.
This assumption is obviously not realistic in large social networks, where nodes are 
generally not aware of the network at large and act mostly within their local social 
neighborhood.\footnote{This relates to the opportunity structure and influences observed 
phenomena like homophily and triadic closure \cite{mcpherson01,kossinets06,jackson07}.}
At the same time, in most social networks \textit{some} edges happen outside the direct social neighborhood \cite{watts03,jackson07}, which suggests the use of mixture models, mixing local and global processes~\cite{kleinberg99,kumar00,jackson07}. From a discrete choice viewpoint, these approaches specify mixed logit models \cite{overgoor19}.

The use of negative sampling does not cleanly transfer to the mixed logit setting \cite{guevara13b}.
Part of our contribution is therefore to introduce a model 
simplification technique for discrete mixture models that we call ``de-mixing'' whereby the individual modes are forced to operate on disjoint sets.
When the modes of a mixed logit model operate over disjoint choice sets,
the resulting model reduces to a collection of individual conditional logit models.
This de-mixing approach thus  opens the door to negative sampling and high performance importance sampling-based approaches to negatives sampling, 
which we separately find to be an important underutilized tool for scaling discrete choice models. De-mixing also circumvents standard challenges with maximizing the likelihoods of mixture models, typically done via an EM algorithm~\cite{train09}.

\xhdr{Importance sampling}
In practice, negative sampling is typically done uniformly, 
a procedure that can be very inefficient when important features are rare within the population.
This inefficiency is particularly pronounced for modeling large-scale social networks,
  as they are frequently driven by activity within one's local social neighborhood \cite{rapoport53,jin01},
  which covers only a small subset of the full node set
  (making ``is a friend of a friend'' a rare feature).
Furthermore, between 50\% and 90\% of new friendship edges form between existing friends-of-friends \cite{backstrom11,leskovec08}. Facebook's People You May Know recommendation system was (as of 2013), built to only rank (and thus recommend) potential friends from existing friends of friends \cite{ugander13}.
This concentration is even more pronounced for relational events, where pairs of individuals can interact repeatedly and often do.
On the Venmo platform, which we analyze in this paper, approximately 73\% of transactions in 2018 happened between people that had interacted before.
In these settings, with the bulk of activity happening within the local neighborhood, uniform negative sampling is not a feasible solution.

To circumvent these issues and enable estimation of choice models on large graphs, we use importance sampling \cite{owen13}, a standard technique for approximate inference, 
to sample non-chosen alternatives non-uniformly. Importance sampling produces consistent estimates
when coupled with an adjustment, based on the probability of being sampled, to the likelihood function.
As a result, we can feasibly fit conditional logit models, both regular and de-mixed, that incorporate activity in the local neighborhood to very large graphs.

\xhdr{Applications}
We first illustrate the value of negative sampling on synthetic data with a known data generating process.
We find that non-uniform importance sampling is especially effective for rare features that commonly drive network formation, 
  where it can reduce the variance of the estimates by an order of magnitude.
As an important inspection, we examine the trade-off between an overall downsampling of the data vs.\ sampling non-chosen alternatives within data points, 
  and generally find that additional data points are more valuable (in terms of mean squared error) than additional negative samples.
Model de-mixing, meanwhile, efficiently translates mixture models into conditional logit models, where negative sampling can be validly applied. 

To illustrate the real-world feasibility of fitting discrete choice models to very large graphs,
 we introduce and analyze a large-scale dataset of public transactions on the Venmo platform.
The data includes 501M public transactions, occurring between April 2012 and July 2018, involving 25M distinct user accounts.
A single conditional logit model on this data results in extreme parameter estimates that are hard to interpret.
However, a de-mixed mixed logit model shows that for local activity, well-known dynamics like reciprocity take place, while activity outside the local neighborhood appears to be primarily driven by preferential attachment.
In this case, the de-mixed model provides significant new insight in the formation dynamics of this real world graph.

By de-mixing models that better approach the availability assumption and by leveraging non-uniform importance sampling,
we find that discrete choice models can be scaled to large-scale relational event data with great potential for impact.
These advances open up their use to diverse data mining applications, including but not limited to link prediction, anomaly detection, and general recommendation systems.

\xhdr{Additional related work}
Discrete choice models have a long history of study in psychology \cite{thurstone27,luce59} and economics \cite{mcfadden77,agresti03,train09}.
The conditional logit functional form has also been used to model network formation both more and less recently,
  including in exponential random graph models (ERGMs) \cite{robins07},
  stochastic actor-oriented models (SAOMs) \cite{snijders10}, 
  relational event models (REMs) \cite{butts08},
  and dynamic network actor models (DyNAMs) \cite{stadtfeld17a}.

The concept of sampling non-chosen alternatives occurs in multiple literatures. 
In statistics and epidemiology, the technique is sometimes called case-control sampling~\cite{langholz95}.
In econometrics, negative sampling is commonly used with conditional logit models~\cite{mcfadden77,manski81,agresti03},
while valid sampling for mixture models is an open problem~\cite{nerella04,guevara13b,vonhaefen18}.
Recent work explores sampling for models of network formation \cite{vu15,overgoor19,lerner19},
  but it is limited to uniform sampling for the conditional logit model.
Our work expands to non-uniform sampling and provides a model simplification approach for mixture models.

In machine learning, negative sampling has found broad adoption for learning representations for text and graph mining \cite{mikolov13,tang15},
though in these contexts the way in which negative sampling changes the likelihood and the resulting representation output (due to model misspecification) is generally ignored.
Thus it is important to emphasize that the change in the likelihood under misspecification has been of great concern to econometricians, where interpretable model parameters is a primary goal, whereas in representation learning the applied utility of the learned representations in downstream prediction tasks is most important, with interpreting the exact representations being of minimal interest.

Lastly, one prior study of Venmo transactions \cite{zhang17} found dense clustering that frequently related to niche uses such as paying rent or settling betting pools. However, they do not model or analyze the factors driving the transactions.

\section{Relational events as choices}

Discrete choice models are commonly employed to model the factors driving
  a choice from a discrete slate of alternatives.
Data points are individual choices $(i, j,C)$,
  indicating that agent $i$ chose item $j$ out of choice set $C$,
  where $C \subseteq U$ is typically a subset of some universe of alternatives $U$.
The most popular class of discrete choice models is the random utility model (RUMs) \cite{manski81},
where each alternative $j \in C$ can be thought of as providing some inherent utility to the agent $i$ making the choice, and the agent makes their choice by maximizing a noisy observation of that utility.

\subsection{Conditional Logit}
\label{sec:cl}

For the conditional logit model, the utility is modeled as a linear function of 
$x_{i,j} \in \mathbb{R}^d$, $j$'s features in the eyes of $i$, plus additive noise $\epsilon_i$:
\begin{eqnarray*}
U_{i,j} = \theta_i^T x_{    i,j}     + \epsilon_i,
\end{eqnarray*}
for some (latent) parameter vector $\theta_i \in \mathbb{R}^d$, which can be fixed or vary across individuals.
Unless explicitly noted otherwise, we assume that the parameter vector $\theta$ is shared across individuals.
When the noise terms are i.i.d.~standard Gumbel, 
  the probability of choosing each alternative 
  is proportional to the exponentiated inherent utility $\theta^T x_{i,j}$ \cite{agresti03}:
\begin{eqnarray}
P_i(j,C) = \frac{\exp{\theta^T x_{i,j}}}{\sum_{\ell \in C} \exp{\theta^T x_{i,\ell}}}.
\label{eq:conditional_logit}
\end{eqnarray}

To see how this framework applies to relational events modeling, 
a relational event initiates a new edge $(i,j)$ in a directed network 
that can be seen as choice where node $i$ has chosen to connect or transact with node $j$, 
as opposed to any other node. The set of candidates for this choice, 
typically the full node set of the network at the time of the event, is then the choice set $C$.
Our goal in fitting a discrete choice model to relational event data
is to understand the potential role of different features $x_{i,j}$, 
which can be node covariates or information about the network structure 
at the time the edge was formed.\footnote{Feature values often change over time, as the network changes. The features $x_{i,j,t}$ of node $j$ at time $t$ are thus always time-indexed, but we generally suppress the $t$ subscripts for notational clarity.}
As a concrete example, preferential attachment \cite{albert99} can be thought of 
as a conditional logit model with the single-element feature vector 
$x_{j,t} = \log d_{j,t}$,
using only the candidate node $j$'s degree $d_{j,t}$ at time $t$.
    Many other network formation dynamics can easily be included in this modeling framework \cite{overgoor19}.

\begin{figure}
    \centering
    \includegraphics[width=0.95\columnwidth]{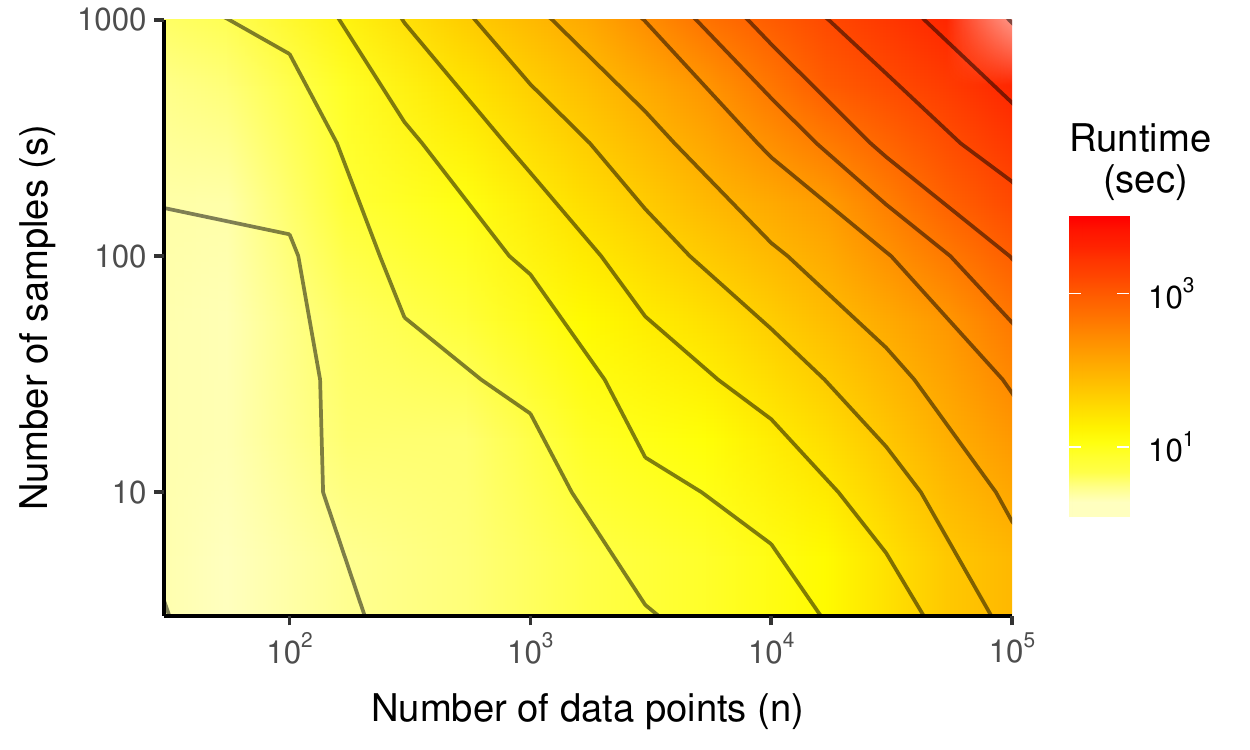}
    \vspace{-5mm}
    \caption{
        The runtime (in seconds) of training a conditional logit model, varying
        the number of choices $n$ and number of alternatives to choose from $s$, is roughly linear in both dimensions. Each point represents the average of 50 runs.
    }
    \label{fig:complexity}
\end{figure}

Conditional logit models with linear utility functions have a log-likelihood that is smooth and convex in the parameters $\theta$, meaning the maximum likelihood parameters can be efficiently estimated using, e.g., gradient descent or BFGS.
The time complexity\footnote{Suppressing  dependence on the precision, number of parameters, and properties of the features.} of fitting a conditional logit model with gradient descent with $n$ training examples and $s$ alternatives per example is then $O(ns)$ \cite{nesterov13}. This dependence of the time complexity on the different problem parameters is illustrated in Figure~\ref{fig:complexity}, where we plot the average runtime\footnote{Runtime  using the BFGS optimizer from \texttt{scipy} with a tolerance of $10^{-8}$.} while varying $n$ and $s$. 
While the runtime is linear in $n s$, different choices of $n$ and $s$ at constant $n s$ can have different bias and accuracy, as we explore in Section~\ref{sec:synth1}.

\subsection{Heterogeneity with known context}
\label{sec:het-known}

The conditional logit, as presented so far, assumes a single parsimonious parameter vector $\theta$ that is shared across all agents and in all contexts, while it's clear that preferences may vary over both agents and contexts.
First, different kinds of \textit{choosers} can have differing preferences.
For example, men and women may have different styles of making friendships, as may students and non-students \cite{currarini10}.
In the limit, individuals can have their own personal preference, though this can only be modeled when one has considerable data for each individual.
On the other hand, the choice \textit{context} could affect preferences.
For example, people make different decisions at different stages in their careers,
  and their choices can depend on their moods or the choice interface \cite{chen16,seshadri19}.

If the chooser type or choice context is \textit{known}, it is relatively straightforward to model heterogenous preferences.
The analyst can fit separate parameters $\theta_i$ for each ``group'' $i$, or,
  if some parameters are hypothesized to be shared across groups,
  interact the context and the context-dependent features.
Because the conditional logit estimates the relative utility of items \textit{within a choice set},
  one cannot simply include the choice context as an interaction term, as one would do with a regression model.
Instead, the interaction can be implemented by one-hot coding each coefficient for each context, and setting the value to 0 when the context feature does not apply \cite{agresti03}.
 This conveniently avoids co-linearity and keep the model identifiable,
  provided that there are enough data points for each context-feature value combination.

\subsection{Mixed Logit for latent context}
\label{sec:het-unknown}
When the choice context is unknown, one can still model heterogeneous choice using a mixed logit approach with latent variables.
In general, a mixed logit model is defined using a continuous probability distribution $f$ over different instances of the parameters $\theta$.
In this work, we will only consider discrete mixtures of $M$ different parameters (called ``modes''), an approach also called a latent class model. The choice probability is then:
\begin{eqnarray}
P_i(j,C) = \sum_{m=1}^M  \pi_m \frac{\exp{\theta^T_m x_j}}{\sum_{\ell \in C_m} \exp{\theta^T_m x_\ell}} \mathbf{1}[ j \in C_m],
\label{eq:mixed_logit}
\end{eqnarray}
where the class weights $\pi_m \ge 0$ model the relative prevalence of each mode and $\sum_{m=1}^M \pi_m = 1$.
Each mode thus has its own preference vector $\theta_m$ and can have its own limited choice set $C_m$ as long as $C_m \subseteq C$.
Maximizing the likelihood of a mixed logit model is harder than for the conditional logit, since the log-likelihood is not convex in general.
A common solution is to use expectation maximization (EM) \cite{train09}.
However, EM can be slow to find even a locally optimal solution, since there often is a trade-off between changing the class weights and changing the feature weights.

A mixed logit model greatly increases the degrees of freedom 
the analyst has to fit a model to their data.
There can be many options for how to specify the choice set $C_m$ of each mode, which could be identical, disjoint, or exhibit arbitrary overlap.
Care should be taken with formulating a hypothesis on how the data was generated,
 and to use model selection and robustness checks while doing inference.

\subsection{Mixed logit and de-mixing}
In modeling relational events, the mixed logit formulation subsumes a number of popular network formation models such as the copy model \cite{kleinberg99,kumar00} 
and the local search model \cite{jackson07,holme12}.
The local search model attempts to address the availability assumption, discussed in the introduction, by mixing two modes where the first mode is restricted to a choice set of friends-of-friends of the chooser $i$, which we denote as $FoF_i$, while the second mode makes unrestricted choices from the full node set $V$. 
The standard local search model assumes very simple utility functions for each of the two modes, letting the difference in choice sets do most of the heavy lifting. Typically the utility function on the unrestricted choice set is constant, meaning that it can really be viewed as a ``garbage collection'' mode\footnote{This approach is not unlike similar strategies that are common with Gaussian mixture models, where a high-variance mode is often used to capture outliers~\cite{tukey60}.}, capturing rare friendship events to non-friends-of-friends without any attempt to model specific mechanisms of such events.
The local search model is thus an attractive approach to addressing the availability assumption, but introduces all of the above disadvantages of mixed logits.
However, in any realistic social network, it's clear that the sets $FoF_i$, for all $i$, will generally be very small subsets of the full node set. This observation motivates the de-mixing approach that we now introduce.

What makes fitting a standard mixed logit model hard is that when a chosen alternative is available in the choice sets $C_m$ of several modes, a given data point can have very similar conditional likelihoods under different modes.
However, when the chosen alternative is only available in the choice set of a single mode, the choice uniquely identifies the mode, collapsing the mixture.
Thus, when the choice sets of the modes are fully disjoint, every data point $j$
has a non-zero likelihood only
for the mode $m$ where $j \in C_m$.
Phrased in the language of Sections~\ref{sec:het-known} and~\ref{sec:het-unknown}, it is as if the choice context is known.
 We call this approach, whereby a mixture model with nearly disjoint choice sets is approximated by a model with forcibly disjoint choice sets, ``de-mixing''.

What does de-mixing give us? When estimating a disjoint mixed logit,
   the class parameters $\pi_m$ then follow directly from the data and
   the preference parameters $\theta_m$ can be found by fitting $m$ individual conditional logit models.
To illustrate,
   the log-likelihood of the model on disjoint choice sets $C_m$ fit to the full data can be written:
\begin{equation*}
  \begin{split}
   l(\theta; \mathcal D) 
      &= \log \prod_{(j,C) \in \mathcal D} \sum_m^M \pi_m \frac{\exp{\theta^T_m x_j}}{\sum_{l \in C_m} \exp{\theta^T_m x_l}}  \mathbf{1}[ j \in C_m] \\
      &= \sum_m^M \log \prod_{(j,C) \in \mathcal D_m} \frac{\exp{\theta^T_m x_j}}{\sum_{l \in C_m} \exp{\theta^T_m x_l}} = \sum_m^M l(\theta_m, \mathcal D_m), 
\  \end{split}
\end{equation*}
where $\mathcal D_m = \{(j,C) \in \mathcal D\ :\ j\in C_m \}$ are the disjoint subset of the data where the choice was made from the mode's reduced choice set.
The mode-level features $\theta_m$ can thus be estimated separately, even in parallel, on the different disjoint subsets of the data. 

Given the greatly simplified likelihood, mixture models with disjoint choice sets are highly attractive.
As we will see in Section~\ref{sec:sampling}, the return to conditional logits re-introduces the validity of negative sampling, at least under the model class, which is essential for scalable inference.
Depending on the analysis context, de-mixing may thus constitute a minor simplification with great practical implications.
For example, in the local search model discussed above, the choice sets $C_1=FoF_i$ and $C_2=V$ are often very nearly disjoint.
By slightly modifying the second choice set to be $C_2=V \setminus FoF_i$, we get a model that is easy to estimate on very large graphs.

\section{Sampling of alternatives}
\label{sec:sampling}

In large real-world graphs, the choice set $C$ of alternatives (all nodes) quickly become prohibitively large.
To make estimating a conditional logit model computationally tractable in such settings,
  one can estimate the same model on \textit{reduced} choice sets of alternatives.
 Specifically, $s-1$ negative/non-chosen examples can be
  sampled to create a (random) reduced
  dataset with smaller choice sets.
For each choice $(j,C)$,
  one forms a smaller random choice set out of the positive choice and the negative samples, 
  $\tilde C \subset C$ with $\lvert \tilde C \rvert =s$, and
  replaces the original choice data with $(j,\tilde C)$.
 
McFadden showed that asymptotically (in the limit of large datasets), estimates on a data set with reduced choice sets generated with importance sampling
  are consistent (in the limit of many data points) for the estimates using complete choice sets,
  provided that the true process is \textit{within the model class} of conditional logit \cite[eq. 40-41]{mcfadden77}.
Phrased another way, this classic econometric result highlights that estimation under negative sampling 
does not offer a consistency guarantee outside the model class.
As a result, outside the model class it is clear that different sampling distributions and different negative sample budgets $s$ can lead to very different parameter estimates.

Within the correct model class,
  to produce estimates that are consistent
  one has to adjust the likelihood function.
Individual examples need to be weighted based on the sampling probability of the reduced choice set $\tilde C$\footnote{Perhaps unintuitively, the chosen element also has to be weighted based on its sampling probability, a point that has lead to confusion in other work \cite{bruch12,jarvis18}.}.

Following \cite{mcfadden77,benakiva85}, for a reduced choice set $\tilde C \subseteq C$ the choice probability is then
\begin{eqnarray}
P_i(j,\tilde C) = \frac{\exp{\big(\theta^T x_j - \log q_{j}}\big)}{\sum_{\ell \in \tilde C} \exp{\big(\theta^T x_\ell - \log q_{\ell}\big)}},
\label{eq:conditional_logit_sample_v2}
\end{eqnarray}
where $q_{x}$ is the probability, under the importance sampling distribution, of including
  element $x$ in the reduced choice set $\tilde C$ independently from other alternatives.

\xhdr{Uniform sampling}
The most straightforward sampling procedure is uniform sampling, which makes every individual alternative equally likely to be chosen.
Under uniform sampling (without replacement\footnote{It is possible, and in some cases more efficient, to sample alternatives with replacement, but this increases the complexity of computing the sampling weights \cite[page 266]{benakiva85}.}),
  the probabilities $q_x$ are all equal and cancel out.
Therefore, no adjustments is necessary to have consistent parameter estimates under uniform sampling.

\xhdr{Stratified sampling}
Uniform negative sampling ignores both the relative frequency and the relative importance of features in the data, and is therefore not an efficient way to estimate the parameters.
  Stratified sampling can be used to create choice sets that are balanced with 
  respect to one or more categorical dimensions. As a relevant example, we can construct strata for friends, friends-of-friends, and other nodes (recall Figure~\ref{fig:diagram}).
Practically, a pre-specified amount of alternatives is sampled for each specific stratum $G$ (uniformly at random within the stratum).
The sampling weights are proportional to the relative sizes of the categories within the full choice sets.
When sampling $s$ alternatives for a specific stratum $G$\footnote{For the stratum of the chosen alternative, $s-1$ samples are drawn so that there are $s$ elements from that stratum in the reduced choice set.}, of which there are $n_G$ elements in the full choice set ($n_G = | G \cap C|$)\footnote{Approaches to deal with the case when $s > n_G$, include leaving the strata to be unequal size, revert to sampling with replacement, or sample the remaining $s-n_G$ alternatives from other strata. We apply the latter in this work. Either way, the sampling weights have to be adjusted accordingly.},
  the sampling weight for an individual item $x$ becomes $q_x = \frac{s}{n_G}$.
Now, Equation \eqref{eq:conditional_logit_sample_v2} can be used to produce consistent estimates of the parameters.
This approach is flexible with respect to the way strata are defined.
For example, multiple dimensions can be crossed and continuous features can be bucketed.

\xhdr{Importance sampling} 
Stratified sampling is a special case of importance sampling.
An importance sampling policy assigns to each element $x$ its own sampling probability $q_x$ \cite[Ch.\ 9]{owen13}.
The estimator is consistent regardless of the choice of the importance sampling weights, but better weights can reduce the variance of the estimator.
More efficient than having balanced strata is to sample elements of $\tilde C$ proportional to the likelihood of being chosen.
In the optimal sampling policy, the relative likelihood of choosing the positive and negative samples should vary as little as possible \cite[page 264]{benakiva85}. 
However, in order to compute these choice probabilities, this hypothetical optimal policy presumes the availability of the very feature weights that we are trying to estimate.
For example, if the actual true feature weight for some binary covariate is $\beta$, then the ideal rate between the sampling probabilities for options with and without that covariate is $\exp(\beta)$.

There are various ways in which proxies for the importance sampling probabilities can be obtained,
  including employing domain knowledge,
  using a parametric model to obtain sampling weights for a non-parametric model,
  sequentially fitting a crude model and then a finer model, 
  and/or to compute unweighted empirical choice probabilities from the data.
We leave exploration of these diverse approaches for future work, focusing on the high relevance of stratified sampling based on friends and friends-of-friends. 

\xhdr{Sampling for Mixed Logit} 
For the general class of mixed logit models, estimates on a (negatively sampled) reduced choice set are \textit{not} consistent for the estimates on the full choice set \cite{guevara13b}.
To correct the bias, the likelihood has to be adjusted with factors $W^1_n,\ldots,W^M_n$ representing the likelihood of sampling the reduced choice set $\tilde C$ when $j$ is chosen for each of the $M$ modes.
Unfortunately, the quantity $W_n^m$ is not easily computed, as it relies on the choice probability of the full choice set(!), rather than the reduced choice set $\tilde C$.
Various heuristic approaches have been proposed to attempt uniform negative sampling  \cite{nerella04,vonhaefen18,guevara13b}.
While empirical results have generally been promising, a full evaluation of these specific approaches has to be left for future work.
The lack of guarantees for fitting mixed logit models on negatively sampled data 
strongly limits their utility for most realistic network data sets. In the de-mixed logit setting, however, the sampling techniques of the conditional logit can be applied to each mode individually.

\section{Applications on Synthetic data}

In this section we illustrate the benefits and applications of sampling non-chosen alternatives for conditional logit models on relational event data.
In the first example we explore the benefits for inference from importance sampling on a data set with a known data generating process. The next example studies how negative sampling
can alter the estimates when the model is misspecified.
Applications to real-world data follow in Section~\ref{sec:venmo}.
For the reproducibility of our results, we've shared the code for these synthetic experiments at \url{https://github.com/janovergoor/choose2grow/}.

\subsection{Synthetic conditional logit experiments}
\label{sec:synth1}

If choice sets are small, it is computationally feasible to consider all non-chosen alternatives when maximizing the likelihood, which will lead to the best parameter estimates given the data.
With relational data, however, the choice sets are typically enormous, necessitating sampling.
Importance sampling is motivated by item features that are unequally distributed in the choice set.
The advantage over uniform sampling is especially pronounced when the number of data points is small, or when it is not feasible to consider many non-chosen alternatives.
To illustrate this advantage, we construct a synthetic data example showing the benefits of stratified sampling, a simple form of importance sampling.

\xhdr{Data generation}
We simulate events happening between members of a fixed population of $N$=5,000 nodes.
To seed a graph, we generate $|E|$=25,000 events by uniformly sampling the nodes $(i,j)$.
We then sequentially generate 160,000 events (data points) where for each event we first sample the ``sending node'' $i$ uniformly at random.
Every possible ``target node'' $j$ for the event then has a utility based on a pre-specified utility function with four principle features taken in two transformations (thus, eight features):
$j$'s in-degree (representing {\it popularity}),
number of prior events from $i$ to $j$ (representing {\it repetition}),
number of prior events from $j$ to $i$ (representing {\it reciprocity}),
and the number of unique friends of friends between $i$ and $j$.
For these four principle features we apply two transformations: we take the log (mapping $\log 0$ to $0$),
as well as an indicator feature for whether the value is non-zero.
We set the parameter values to $(0.5, 2, 2, 2)$ for the log features, and $(1, 4, 4, 4)$ for the indicator features.

\xhdr{Varying $n$}
We construct data sets of incrementally increasing size (so that the smaller data set is contained in the larger data sets), and for each total of $n$ data points, we consider $s-1$ negative samples for $s=24$ under both uniform and stratified sampling.
For stratified sampling, $s/3$ negative samples are taken from each of three groups: friends (have interacted with before), friends-of-friends (friends have interacted with before), and the rest.
For every incremental dataset, we fit a conditional logit model with the same specification as the utility function we used to generate the data.

\begin{figure}
        \centering
        \includegraphics[width=0.95\columnwidth]{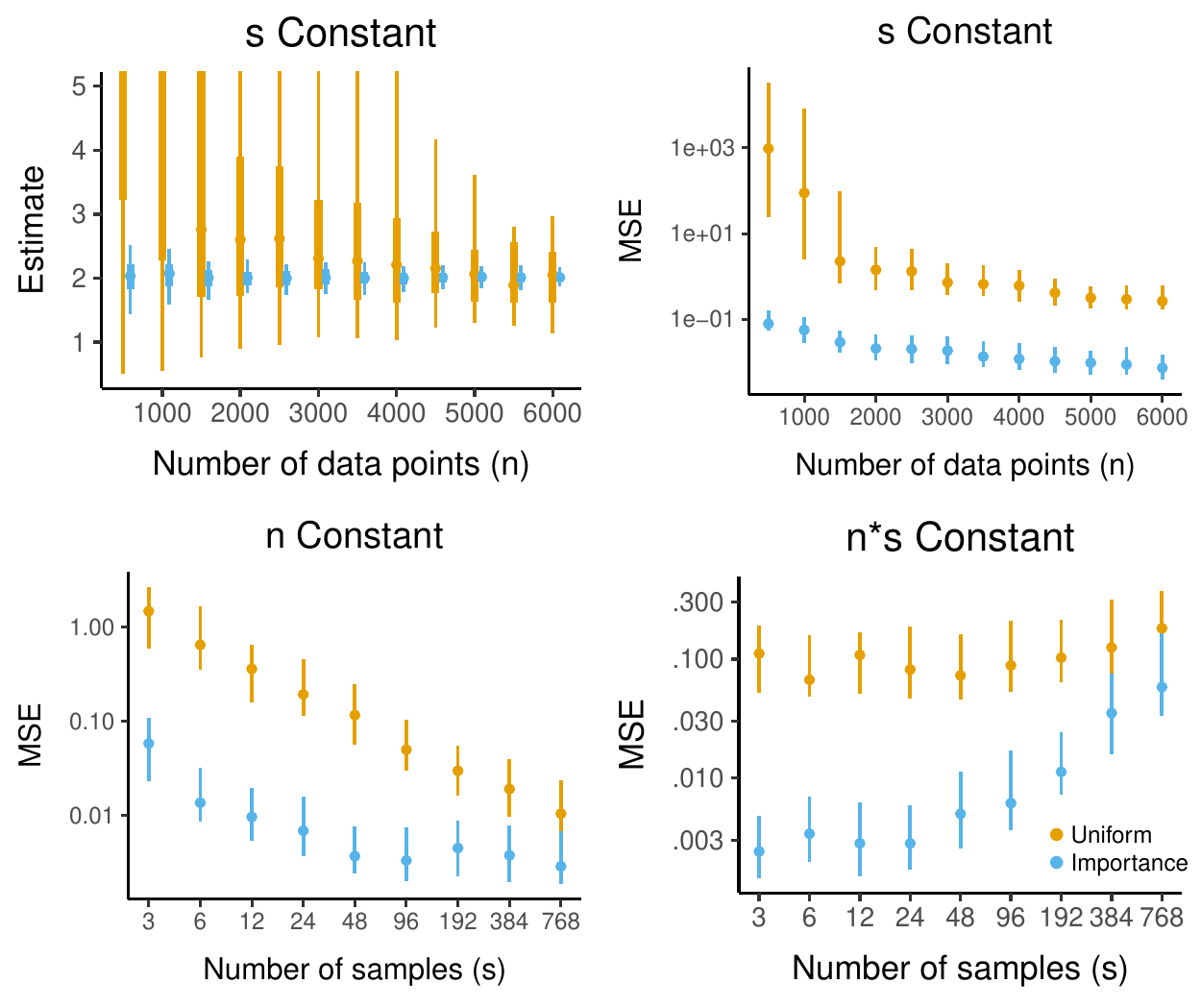}
  \vspace{-4mm}
  \caption{
    Top row: The range of estimates (left) and MSE of the point estimates (right) for the \textit{reciprocate} feature (true value $\theta_{\text{rec}}$=2) when varying $n$.
    Bars represent 5/95 and 25/75 inter-quantile ranges of 100 estimates.
    Importance sampling has an MSE that's an order of magnitude smaller than uniform sampling.
    Bottom row: the range of MSEs, when varying $s$, keeping $n$ constant at $n$=10,000 (left) and with a fixed budged for $ns$=480,000 (right).
    Importance sampling achieves high-precision estimates even for low values of $s$.
    With a fixed budget on $ns$, MSE is lowest for small $s$.
  }
  \label{fig:synth_1}
\end{figure}

We focus our discussion on one representative feature, \textit{reciprocity}, as an example of a feature that benefits from oversampling from the immediate social neighborhood.
In Figure~\ref{fig:synth_1} we plot the ranges of the estimates (top-left) and mean squared error of the estimates (top-right).
As expected, the error decreases as the number of data points increases for both negative sampling methods.
Also, as $n$ increases, the MLE converges to the true value of $\theta_{\text{rec}}$=2.
For low values of $n$ and $s$, there are simply not enough examples of non-chosen alternatives that can be reciprocated with, so non-uniform sampling is much more effective.
The improvement in the error from using stratified sampling (over uniform sampling) is approximately an order of magnitude at these values of $n$.
The other features that are also related to the chooser's social neighborhood similarly benefit from stratified sampling (not shown).
The benefit of importance sampling is most pronounced for features that are unequally distributed in the data of alternatives.

\xhdr{Varying $s$}
Next, we fit the same model, but keep the number of training examples constant at $n$=10,000 and vary $s$.
The range of mean squared errors for the reciprocity parameter are again shown in the bottom-left panel of Figure~\ref{fig:synth_1}.
Uniform negative sampling benefits from increases in $s$.
Importance sampling achieves high-precision estimates even for low values of $s$, because each data points has a minimum number of high-utility alternatives.
For higher value of $s$, importance sampling achieves the same MSE as the estimate of the full choice set without negative sampling.

\xhdr{Varying $n$ and $s$ under a budget}
Increasing $s$ improves the estimates for uniform sampling, but doing so linearly increases the volume of the training data and the runtime of the optimization procedure, as indicated by the discussion of time-complexity in Section~\ref{sec:cl}.
Therefore, paralleling an analysis of SAOM~\cite{snijders10} in \cite{lerner19}, we ask whether increasing $n$ or $s$ is preferred at a given runtime budget.
We study a fixed budget of $ns$=480,000, so that as we increase $s$, we decrease $n$.
We display the resulting estimates for $\theta_{\text{rec}}$ in the bottom-right panel of Figure~\ref{fig:synth_1}.
For both uniform and non-uniform sampling, smaller $s$ perform better. However, for uniform sampling this effect is small as the MSE is relatively uniform. Meanwhile, for stratified sampling we observe significantly more precise estimates at lower values of $s$, up until $s=24$.
We emphasize that this specific inflection point ($s=24, n=20,000$) does not provide a general recommendation for practitioners about how to optimally trade off $n$ and $s$, but rather, we provide this analysis as a demonstration that practitioners should actively consider this trade-off when training choice models of relational data.

\subsection{Model misspecification}
\label{sec:synth2}

As mentioned in the introduction, McFadden's celebrated result on consistency under negative sampling assumes that the data was generated from the correct model class.
In this section, we examine negative sampling when used outside the model class.

\xhdr{Data generation} 
We use a similar data generation process as in Section~\ref{sec:synth1}.
We generate a single event graph with $N=$5,000 nodes and seed it with uniformly sampled pre-events.
We then generate 80,000 choices from a 2-mode mixed logit with disjoint choice sets (a de-mixed model).
The first (``local'') mode considers only friends and friends-of-friends,
  and has the feature vector $\theta = (0.5, 1, 1, 1, 0.5, 1, 1, 1)$. Each mode has the same eight feature as in Section~\ref{sec:synth1}.
The second mode considers all {\it other} nodes (recall it is a de-mixed model),
and has the feature vector $\theta = (1, 0, 0, 0, 1, 0, 0, 0)$, which effectively gives utility only to degree.
The local mode has weight $\pi=0.75$ in the mixture model.

\begin{figure}
    \centering
    \includegraphics[width=0.95\columnwidth]{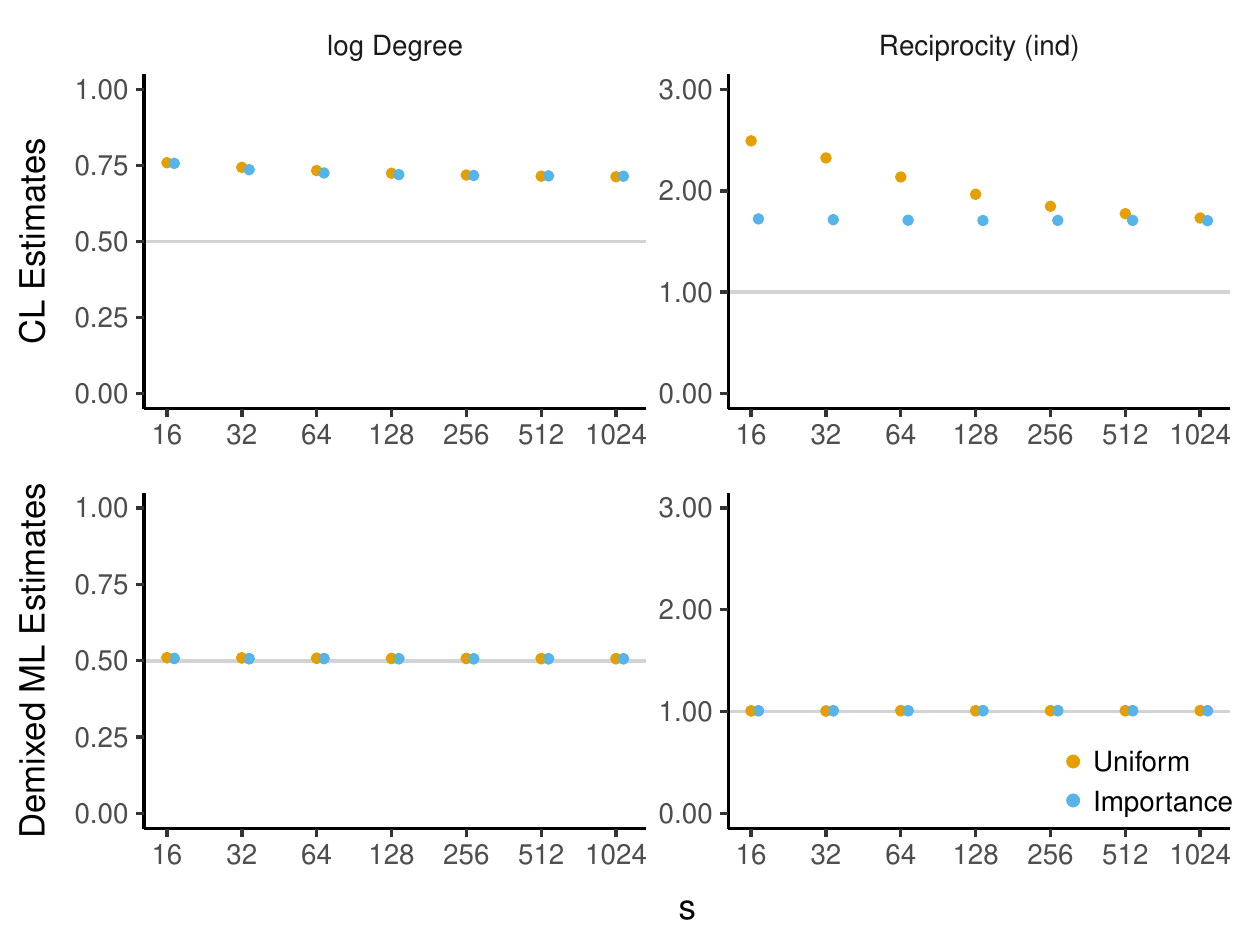}
    \vspace{-6mm}
    \caption{
      Estimates for the \textit{log degree} (left) and \textit{reciprocate} (right) features under 2-mode mixed-logit data, fit with a conditional logit model (top) and a mixed-logit model with a local neighborhood mode (bottom).
      Bars (not visible) represent 5/95 and 25/75 inter-quantile ranges of 100 estimates.
      The misspecified CL estimates vary as a function of $s$ and the sampling methods.
      The correctly specified ML estimates are well-behaved when negatively sampled.
    }
    \label{fig:synth_2}
\end{figure}

\xhdr{Fitting a misspecified conditional logit}
First, we fit a single conditional logit model to this data, with correctly specified features but only as a single mode instead of two.
The resulting estimates under different values of $s$ are shown in Figure \ref{fig:synth_2} (top row).
We focus on two features:
  \textit{log degree} and the indicator function for \textit{reciprocity}.
The range of estimates for both features is practically in-between the true values of the parameter for the two modes (0.5 and 1 for \textit{log degree} and 4 and 0 for \textit{reciprocity}).
Worse, the estimates are not stable across different values of $s$.
Uniform sampling especially gives very different estimates for low values of $s$.
Only when $s$ approaches the full choice set do the two different sampling methods agree.
Caution should always be taken when fitting a conditional logit with negative sampling, and it is important to verify that different meta-decisions (such as $n$, $s$, and the sampling policy) give stable estimates.
If not, this stability analysis can be a diagnostic signal that the data was generated outside the model class.

\xhdr{De-mixed model fit}
Next, we fit the correctly specified model class to the data: a mixed logit with disjoint choice sets.
In practice, this amounts to separating the data points into the different modes based on the features of the chosen example, and fitting a conditional logit to the alternatives from the same mode.
The class probability $\pi_m$ follow directly from the data, in this case $\hat{\pi}_1=0.743$.
In Figure \ref{fig:synth_2} (bottom row), we present the estimates for the ``local'' mode.
The estimates are highly accurate.
Also, the two sampling policies gives relatively stable estimates under different values of $s$.
This is not surprising, as the de-mixed modeling approach places the model within the conditional logit model class.

\section{Empirical analysis of Venmo Data}
\label{sec:venmo}

Venmo is a mobile payment platform that supports person-to-person financial transactions.
Since 2012 Venmo has seen a rapid growth in its transaction volume, both through growth in its active user base and in payments per user.
As compared to other mobile transaction platforms, Venmo usage occurs more within social groups \cite{zhang17}.
Therefore, the structure of the Venmo payment network is likely correlated with the structure of the social interactions of its users, and may exhibit similar structural phenomena.

The Venmo graph records transactions that can occur multiple times between the same two people, similar to instant messaging networks \cite{leskovec08b}.
As of July 2018, Venmo published all of its public transactions through their API,
which allowed anyone to request all the public transactions that occurred within a specified timeframe.
Transactions up until that point were public by default if both of the accounts have their privacy settings set to ``public by default''.

\begin{figure}
    \centering
    \includegraphics[width=\columnwidth]{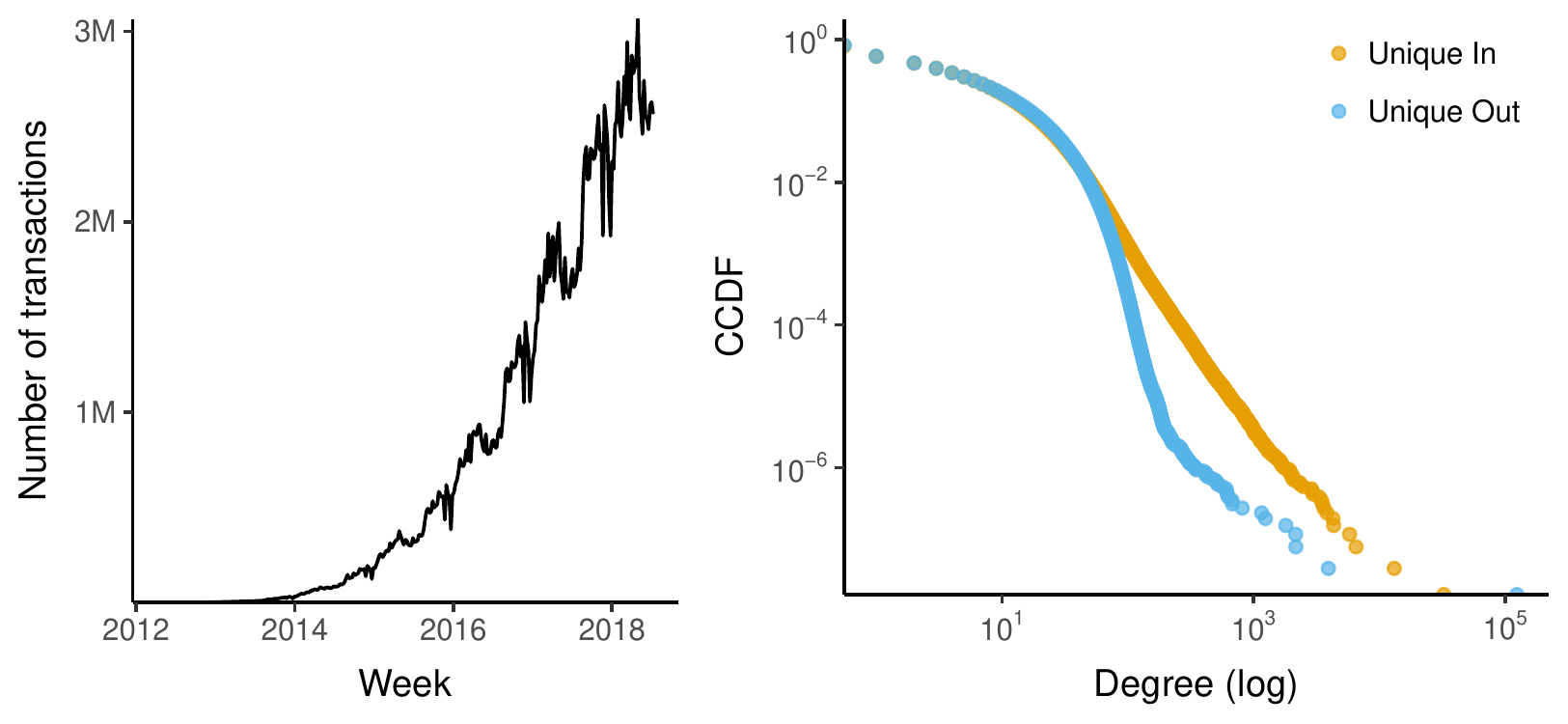}
    \vspace{-6mm}
    \caption{
       The number of transactions per week (left)
         and the complementary cumulative degree distribution (right) of the Venmo transaction dataset.
       Degree is counted as the number of unique transaction alters.
        }
    \label{fig:venmo}
\end{figure}

We retrieved all public transactions between April 12, 2012 and July 17, 2018,
collected under an IRB-approved research protocol.
The data includes 25M users, out of a total of 39M (derived from the incremental user IDs).
The transaction IDs imply that there were 501M public transactions out of a total of approximately 
1B transactions.
Among the transacting actors are individual user accounts as well as companies 
and organizations identifiable through their user name.
Venmo grew considerably during the time span of the dataset, and
we can see that the number of users and transactions grow in shown in Figure~\ref{fig:venmo} (left), where it is also easy to see seasonal effects. 
The overall degree distribution, computed as 
the number of unique accounts transacted with, is shown in Figure~\ref{fig:venmo} (right).
There are not so many accounts sending large amounts of transactions, with the cash-back company \textit{Ibotta} as a single previously-documented exception~\cite{zhang17}.
Transactions primarily re-occur within a local social neighborhood: 
in 2018, 73\% of transactions happened between people that had 
transacted previously and another 6.8\% happened between friends-of-friends.

\subsection{Model-based analysis}

To understand the dynamics in the transaction data,
  we fit both a conditional logit and a de-mixed mixed logit model to the Venmo data.
Following standard practice in relational event modeling \cite{butts08,lerner19}, we construct the following features from the event data:

\begin{itemize}
  \item $|F_j|$ -- the number of unique nodes interacting with node $j$,
  \item $\#_{\rightarrow j}$ -- the times $j$ received money, proxy for $j$'s popularity,
  \item $\#_{i \rightarrow j}$ -- the times $i$ sent $j$ money before,  proxy for repetition,
  \item $T_{\rightarrow j}$ -- the time since $j$ received money (in seconds),
  \item $T_{j \rightarrow}$ -- the time since $j$ sent money,
  \item $T_{i \rightarrow j}$ -- the time since $i$ sent $j$ money,  proxy for repetition,
  \item $T_{j \rightarrow i}$ -- the time since $j$ sent $i$ money,  proxy for reciprocity,
  \item $|FoF_{ij}|$ -- the number of unique nodes forming length-2 forward paths from $i$ to $j$.
\end{itemize}

Each temporal feature is transformed with $\log T^{-1}$
  and each counting measure is log-transformed (where $\log 0 = 0$)
  and has an indicator feature (indicating if that the value is non-zero).
We include the time feature to focus on recency, rather than volume.
 However, we do keep an indicator function for $j$ receiving any message and $j$ receiving a message from $i$.
 
We sampled 1\% the public Venmo transactions between 2018-06-26 and 2018-07-17 for analysis.
We fit both a conditional logit model and a de-mixed mixed logit model to illustrate their differences.
For the conditional logit model, we constructed the data using stratified negative sampling,
    with strata for friends, friends-of-friends, and the rest ($100$ nodes/stratum).
For the mixed logit model, there are two modes that can be estimated separately.
  The first ``local'' mode considers only friends and friends-of-friends of the choosing node as possible candidates.
  Since friends are still much more likely to be chosen than friends-of-friends, we also apply stratified negative sampling with two strata for friends and friends-of-friends  ($100$ nodes/stratum here).
  The second ``rest'' mode considers all other nodes as possible candidates.
  Negative examples for this mode are sampled uniformly.

\subsection{Results}

The resulting parameter estimates are displayed in Table \ref{tab:venmo_res}.
For each coefficient, the standard errors are also provided.

\xhdr{Conditional logit} 
The estimates for the conditional logit model are dominated by the indicator feature for $i$ having transacted with $j$ before.
The parameter estimate of 13.443 would imply a factor of 689,000x ($\approx \exp 13.443$) more likely to be chosen.
This value is directly related to the relative size of the non-local node set, and thus hard to interpret.
The number of friends-of-friends is also a strong factor related to the local neighborhood.
For the rest, the recency of $j$ receiving any money if a major factor.
The positive estimate for the feature representing the time since a transaction from $j$ to $i$ is soft evidence for a version of reciprocity. 
Net of these other features, the negative coefficient for $\log |F_j|$ is consistent with \textit{negative} ``preferential attachment''~\cite{albert99}.

\xhdr{De-mixed mixed logit}
Because the modes of the de-mixed mixed logit are defined over disjoint choice sets,
  the class probability estimates follow directly from counts of the data.
For this month of data, 82.9\% of transactions happened within the local neighborhood.
A major difference compared to a single-mode model is that the coefficient for popularity ``split'' into a negative component for local ties and a positive component for non-local ties.
When a node is already known, its popularity is less important. But when it is not known, popularity is a major predictor of connecting.
The other coefficients show minor differences with respect to the single-mode model.
The coefficient for ``is friend'' is still very strong in the local mode, but orders of magnitude less so.
Reciprocity is strong, with both $j$ sending and $j$ sending to $i$ being important features.

\begin{figure}
    \centering
    \includegraphics[width=0.95\columnwidth]{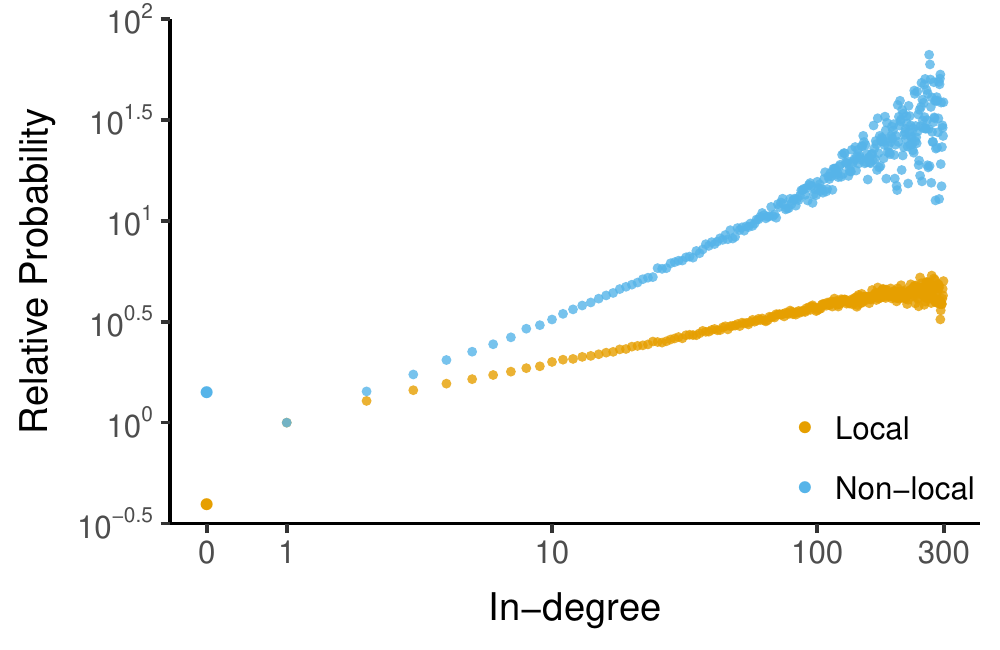}
    \vspace{-3mm}
    \caption{
        Non-parametric estimates of the relative value (compared to degree 1) of in-degree
        for receiving a transaction on Venmo.
        Degree matters less for nodes in the local neighborhood and is valued super-linearly for other nodes.
    }
    \label{fig:non-parametric}
\end{figure}

\xhdr{Non-parametric estimates}
To further illustrate the difference between the two modes,
  we fit another de-mixed mixed logit model with the in-degree (number of transactions) as a (non-parametric) feature,
  so there is just an indicator feature for every individual in-degree.
We plot the estimated relative probabilities of choosing a node $j$ with degree $k_j$,
  as compared to choosing a node with degree 1, in Figure \ref{fig:non-parametric}.
As in the full model results, in-degree matters less for nodes in local neighborhoods than it does for nodes that are further out.
For nodes outside the neighborhood, the in-degree has a super-linear relationship with utility.
These high-degree nodes probably represent small businesses or organizations that receive incoming transactions from many different actors \cite{zhang17}.

\begin{table}[tbp]
  \caption{Logit model fits for Venmo transaction data.
  The conditional logit model is estimated using stratified negative sampling.
  The mixed logit model has two modes with disjoint choice sets:
  one for local nodes (friends and friends-of-friends) and one for the rest.
 }
  \label{tab:venmo_res}
\centering 
\small
\tabcolsep=0.2cm
\begin{tabular}{lrrr} 
\toprule
 & \multicolumn{1}{c}{Conditional} & \multicolumn{2}{c}{Mixed Logit} \\
\cmidrule{3-4} 
&  \multicolumn{1}{c}{Logit} & \multicolumn{1}{c}{Local} & \multicolumn{1}{c}{Rest} \\
\midrule
log $|F_j     |$                           &  -0.032* (0.001)  & -0.377* (0.001)                    &    0.573*  (0.002)  \\
log $T^{-1}_{\rightarrow j}$               &  -0.192* (0.002)  &  0.357* (0.000)                    &    0.413*  (0.001)  \\
log $T^{-1}_{j \rightarrow}$               &   0.381* (0.000)  & -0.015* (0.000)                    &   -0.056*  (0.001)  \\
log $T^{-1}_{i \rightarrow j}$             &  -0.002* (0.000)  &  0.085* (0.001)                    &                    \\
log $T^{-1}_{j \rightarrow i}$             &   0.102* (0.000)  &  0.375* (0.000)                    &                    \\
log $|FoF_{ij}|$                           &   0.390* (0.000)  &  0.451* (0.002)                    &                    \\
$\mathbbm{1}[\#_{\rightarrow j}   > 0]$    &  -0.422* (0.005)  & -1.115* (0.008)                    &  -1.290*  (0.007)  \\
$\mathbbm{1}[|F_j     |  > 0]$             &  -0.624* (0.006)  &  0.000\textcolor{white}{*} (0.032) &  -1.898*  (0.007)  \\
$\mathbbm{1}[\#_{i \rightarrow j} > 0]$    &  13.443* (0.003)  &  3.008* (0.003)                    &                    \\
$\mathbbm{1}[|FoF_{ij}|  > 0]$             &   1.046* (0.002)  &  0.336* (0.002)                    &                    \\
\midrule 
$\pi_m$ &      &  0.829  &   0.171 \\
Observations   &  129,098  &   106,540  &   21,973    \\
\midrule \\[-2.8ex]
\textit{Note:}  & \multicolumn{3}{r}{*p$<$0.01} \\
\bottomrule
\end{tabular} 
\end{table}

\section{Discussion }

Understanding the factors driving choice behavior has many applications in the study of social networks and other relational data.
Discrete choice models provide a principled way to model the multiple factors driving these decisions.
Model-based analysis allows one to analyze, for example, how a platform change 
affects user behavior in a randomized experiment.
Having a more detailed model of behavior also enables anomaly detection by identifying unlikely behaviors or sequences of behaviors.
In this work, we have proposed methods to fit choice models to large-scale relational event data.
De-mixing decomposes the mixed logit into multiple conditional logit models that are easy to estimate.
Importance sampling allows for the estimation of rare features.
Both methods fit well with the insight that most activity in social networks is inherently local.

We briefly point to several open research questions.
First, a more formal theory of when it is better to increase the number of data points $n$ or the number of choice alternatives $s$ is needed (``to sample or to negatively sample?'').
Decisions along this tradeoff likely depend on the model class, distribution of the features, and the variance of the ``true'' covariates.
We also note the need to further work out how varied choice models, e.g.~the nested logit model or context-dependent utility model~\cite{seshadri19}, can be employed for modeling network formation, and how negative sampling works for those models.

Even though the de-mixed mixed logit model has many applications,
  many other applications require the ability to consider overlapping choice sets.
For example, in some applications, different transaction types can apply to overlapping sets of targets.
For these applications to be practically feasible, it is important to be able to fit mixed logit models on reduced choice data.
Further studies of the sampling adjustment for the mixed logit would be welcome.
Finally, there is potential work in finding bridges between choice models and other data mining applications such as sequence embedding models and recommender systems.

\xhdr{Acknowledgements} We thank Austin Benson, Lodewijk Gelauff, J{\"u}rgen Lerner, John Levi Martin, and Arjun Seshadri for helpful comments and discussions.
This work is funded in part by support from the Stanford Thailand Research Consortium and a Young Investigator Award from the Army Research Office (73348-NS-YIP). Data collection approved under Stanford protocol IRB-46422.

\bibliographystyle{ACM-Reference-Format}
\balance
\bibliography{venmo}


\begin{thebibliography}{50}


\ifx \showCODEN    \undefined \def \showCODEN     #1{\unskip}     \fi
\ifx \showDOI      \undefined \def \showDOI       #1{#1}\fi
\ifx \showISBNx    \undefined \def \showISBNx     #1{\unskip}     \fi
\ifx \showISBNxiii \undefined \def \showISBNxiii  #1{\unskip}     \fi
\ifx \showISSN     \undefined \def \showISSN      #1{\unskip}     \fi
\ifx \showLCCN     \undefined \def \showLCCN      #1{\unskip}     \fi
\ifx \shownote     \undefined \def \shownote      #1{#1}          \fi
\ifx \showarticletitle \undefined \def \showarticletitle #1{#1}   \fi
\ifx \showURL      \undefined \def \showURL       {\relax}        \fi
\providecommand\bibfield[2]{#2}
\providecommand\bibinfo[2]{#2}
\providecommand\natexlab[1]{#1}
\providecommand\showeprint[2][]{arXiv:#2}

\bibitem[\protect\citeauthoryear{Agrawal, Imieli{\'n}ski, and Swami}{Agrawal
  et~al\mbox{.}}{1993}]%
        {agrawal93}
\bibfield{author}{\bibinfo{person}{Rakesh Agrawal}, \bibinfo{person}{Tomasz
  Imieli{\'n}ski}, {and} \bibinfo{person}{Arun Swami}.}
  \bibinfo{year}{1993}\natexlab{}.
\newblock \showarticletitle{Mining association rules between sets of items in
  large databases}. In \bibinfo{booktitle}{\emph{ICMD}}.
  \bibinfo{publisher}{ACM}, \bibinfo{pages}{207--216}.
\newblock


\bibitem[\protect\citeauthoryear{Agresti}{Agresti}{2003}]%
        {agresti03}
\bibfield{author}{\bibinfo{person}{Alan Agresti}.}
  \bibinfo{year}{2003}\natexlab{}.
\newblock \bibinfo{booktitle}{\emph{Categorical data analysis}}.
  Vol.~\bibinfo{volume}{482}.
\newblock \bibinfo{publisher}{John Wiley \& Sons}.
\newblock


\bibitem[\protect\citeauthoryear{Akoglu, Tong, and Koutra}{Akoglu
  et~al\mbox{.}}{2015}]%
        {akoglu15}
\bibfield{author}{\bibinfo{person}{Leman Akoglu}, \bibinfo{person}{Hanghang
  Tong}, {and} \bibinfo{person}{Danai Koutra}.}
  \bibinfo{year}{2015}\natexlab{}.
\newblock \showarticletitle{Graph based anomaly detection and description: a
  survey}.
\newblock \bibinfo{journal}{\emph{Data mining and knowledge discovery}}
  \bibinfo{volume}{29}, \bibinfo{number}{3}, \bibinfo{pages}{626--688}.
\newblock


\bibitem[\protect\citeauthoryear{Albert and Barab{\'a}si}{Albert and
  Barab{\'a}si}{1999}]%
        {albert99}
\bibfield{author}{\bibinfo{person}{R{\'e}ka Albert} {and}
  \bibinfo{person}{Albert-L{\'a}szl{\'o} Barab{\'a}si}.}
  \bibinfo{year}{1999}\natexlab{}.
\newblock \showarticletitle{{Emergence of scaling in random networks}}.
\newblock \bibinfo{journal}{\emph{Science}} \bibinfo{volume}{286},
  \bibinfo{number}{5439}, \bibinfo{pages}{509--512}.
\newblock


\bibitem[\protect\citeauthoryear{Backstrom and Leskovec}{Backstrom and
  Leskovec}{2011}]%
        {backstrom11}
\bibfield{author}{\bibinfo{person}{Lars Backstrom} {and} \bibinfo{person}{Jure
  Leskovec}.} \bibinfo{year}{2011}\natexlab{}.
\newblock \showarticletitle{Supervised random walks: predicting and
  recommending links in social networks}. In \bibinfo{booktitle}{\emph{WSDM}}.
  \bibinfo{publisher}{ACM}, \bibinfo{pages}{635--644}.
\newblock


\bibitem[\protect\citeauthoryear{Ben-Akiva, Lerman, and Lerman}{Ben-Akiva
  et~al\mbox{.}}{1985}]%
        {benakiva85}
\bibfield{author}{\bibinfo{person}{Moshe~E Ben-Akiva},
  \bibinfo{person}{Steven~R Lerman}, {and} \bibinfo{person}{Steven~R Lerman}.}
  \bibinfo{year}{1985}\natexlab{}.
\newblock \bibinfo{booktitle}{\emph{Discrete choice analysis: theory and
  application to travel demand}}. Vol.~\bibinfo{volume}{9}.
\newblock \bibinfo{publisher}{MIT press}.
\newblock


\bibitem[\protect\citeauthoryear{Benson, Abebe, Schaub, Jadbabaie, and
  Kleinberg}{Benson et~al\mbox{.}}{2018a}]%
        {benson18b}
\bibfield{author}{\bibinfo{person}{Austin~R Benson}, \bibinfo{person}{Rediet
  Abebe}, \bibinfo{person}{Michael~T Schaub}, \bibinfo{person}{Ali Jadbabaie},
  {and} \bibinfo{person}{Jon Kleinberg}.} \bibinfo{year}{2018}\natexlab{a}.
\newblock \showarticletitle{Simplicial closure and higher-order link
  prediction}.
\newblock \bibinfo{journal}{\emph{PNAS}} \bibinfo{volume}{115},
  \bibinfo{number}{48}, \bibinfo{pages}{E11221--E11230}.
\newblock


\bibitem[\protect\citeauthoryear{Benson, Kumar, and Tomkins}{Benson
  et~al\mbox{.}}{2018b}]%
        {benson18}
\bibfield{author}{\bibinfo{person}{Austin~R Benson}, \bibinfo{person}{Ravi
  Kumar}, {and} \bibinfo{person}{Andrew Tomkins}.}
  \bibinfo{year}{2018}\natexlab{b}.
\newblock \showarticletitle{A Discrete Choice Model for Subset Selection}. In
  \bibinfo{booktitle}{\emph{WSDM}}. \bibinfo{publisher}{ACM},
  \bibinfo{pages}{37--45}.
\newblock


\bibitem[\protect\citeauthoryear{Bruch and Mare}{Bruch and Mare}{2012}]%
        {bruch12}
\bibfield{author}{\bibinfo{person}{Elizabeth~E Bruch} {and}
  \bibinfo{person}{Robert~D Mare}.} \bibinfo{year}{2012}\natexlab{}.
\newblock \showarticletitle{Methodological issues in the analysis of
  residential preferences, residential mobility, and neighborhood change}.
\newblock \bibinfo{journal}{\emph{Sociological methodology}}
  \bibinfo{volume}{42}, \bibinfo{number}{1}, \bibinfo{pages}{103--154}.
\newblock


\bibitem[\protect\citeauthoryear{Butts}{Butts}{2008}]%
        {butts08}
\bibfield{author}{\bibinfo{person}{Carter~T Butts}.}
  \bibinfo{year}{2008}\natexlab{}.
\newblock \showarticletitle{{A Relational Event Framework for Social Action}}.
\newblock \bibinfo{journal}{\emph{Sociological Methodology}}
  \bibinfo{volume}{38}, \bibinfo{number}{1}, \bibinfo{pages}{155--200}.
\newblock


\bibitem[\protect\citeauthoryear{Chen and Joachims}{Chen and Joachims}{2016}]%
        {chen16}
\bibfield{author}{\bibinfo{person}{Shuo Chen} {and} \bibinfo{person}{Thorsten
  Joachims}.} \bibinfo{year}{2016}\natexlab{}.
\newblock \showarticletitle{Predicting matchups and preferences in context}. In
  \bibinfo{booktitle}{\emph{KDD}}. \bibinfo{publisher}{ACM},
  \bibinfo{pages}{775--784}.
\newblock


\bibitem[\protect\citeauthoryear{Currarini, Jackson, and Pin}{Currarini
  et~al\mbox{.}}{2010}]%
        {currarini10}
\bibfield{author}{\bibinfo{person}{Sergio Currarini},
  \bibinfo{person}{Matthew~O Jackson}, {and} \bibinfo{person}{Paolo Pin}.}
  \bibinfo{year}{2010}\natexlab{}.
\newblock \showarticletitle{Identifying the roles of race-based choice and
  chance in high school friendship network formation}.
\newblock \bibinfo{journal}{\emph{PNAS}} \bibinfo{volume}{107},
  \bibinfo{number}{11}, \bibinfo{pages}{4857--4861}.
\newblock


\bibitem[\protect\citeauthoryear{Ghasemian, Hosseinmardi, Galstyan, Airoldi,
  and Clauset}{Ghasemian et~al\mbox{.}}{2019}]%
        {ghasemian19}
\bibfield{author}{\bibinfo{person}{Amir Ghasemian}, \bibinfo{person}{Homa
  Hosseinmardi}, \bibinfo{person}{Aram Galstyan}, \bibinfo{person}{Edoardo~M
  Airoldi}, {and} \bibinfo{person}{Aaron Clauset}.}
  \bibinfo{year}{2019}\natexlab{}.
\newblock \showarticletitle{Stacking Models for Nearly Optimal Link Prediction
  in Complex Networks}.
\newblock \bibinfo{journal}{\emph{arXiv preprint arXiv:1909.07578}}.
\newblock


\bibitem[\protect\citeauthoryear{Guevara and Ben-Akiva}{Guevara and
  Ben-Akiva}{2013}]%
        {guevara13b}
\bibfield{author}{\bibinfo{person}{C~Angelo Guevara} {and}
  \bibinfo{person}{Moshe~E Ben-Akiva}.} \bibinfo{year}{2013}\natexlab{}.
\newblock \showarticletitle{{Sampling of alternatives in Logit Mixture
  models}}.
\newblock \bibinfo{journal}{\emph{Transportation Research Part B}}
  \bibinfo{volume}{58}, \bibinfo{number}{C}, \bibinfo{pages}{185--198}.
\newblock


\bibitem[\protect\citeauthoryear{Holme and Saram{\"a}ki}{Holme and
  Saram{\"a}ki}{2012}]%
        {holme12}
\bibfield{author}{\bibinfo{person}{Petter Holme} {and} \bibinfo{person}{Jari
  Saram{\"a}ki}.} \bibinfo{year}{2012}\natexlab{}.
\newblock \showarticletitle{Temporal networks}.
\newblock \bibinfo{journal}{\emph{Physics Reports}} \bibinfo{volume}{519},
  \bibinfo{number}{3}, \bibinfo{pages}{97--125}.
\newblock


\bibitem[\protect\citeauthoryear{Jackson and Rogers}{Jackson and
  Rogers}{2007}]%
        {jackson07}
\bibfield{author}{\bibinfo{person}{Matthew~O Jackson} {and}
  \bibinfo{person}{Brian~W Rogers}.} \bibinfo{year}{2007}\natexlab{}.
\newblock \showarticletitle{{Meeting Strangers and Friends of Friends: How
  Random Are Social Networks?}}
\newblock \bibinfo{journal}{\emph{AER}} \bibinfo{volume}{97},
  \bibinfo{number}{3}, \bibinfo{pages}{890--915}.
\newblock


\bibitem[\protect\citeauthoryear{Jarvis}{Jarvis}{2018}]%
        {jarvis18}
\bibfield{author}{\bibinfo{person}{Benjamin~F Jarvis}.}
  \bibinfo{year}{2018}\natexlab{}.
\newblock \showarticletitle{Estimating Multinomial Logit Models with Samples of
  Alternatives}.
\newblock \bibinfo{journal}{\emph{Sociological Methodology}}
  \bibinfo{volume}{49}, \bibinfo{number}{1}, \bibinfo{pages}{341--348}.
\newblock


\bibitem[\protect\citeauthoryear{Jin, Girvan, and Newman}{Jin
  et~al\mbox{.}}{2001}]%
        {jin01}
\bibfield{author}{\bibinfo{person}{Emily~M Jin}, \bibinfo{person}{Michelle
  Girvan}, {and} \bibinfo{person}{Mark~EJ Newman}.}
  \bibinfo{year}{2001}\natexlab{}.
\newblock \showarticletitle{Structure of growing social networks}.
\newblock \bibinfo{journal}{\emph{Physical Review E}} \bibinfo{volume}{64},
  \bibinfo{number}{4}, \bibinfo{pages}{046132}.
\newblock


\bibitem[\protect\citeauthoryear{Kleinberg, Kumar, Raghavan, Rajagopalan, and
  Tomkins}{Kleinberg et~al\mbox{.}}{1999}]%
        {kleinberg99}
\bibfield{author}{\bibinfo{person}{Jon~M Kleinberg}, \bibinfo{person}{Ravi
  Kumar}, \bibinfo{person}{Prabhakar Raghavan}, \bibinfo{person}{Sridhar
  Rajagopalan}, {and} \bibinfo{person}{Andrew~S Tomkins}.}
  \bibinfo{year}{1999}\natexlab{}.
\newblock \showarticletitle{The web as a graph: measurements, models, and
  methods}. In \bibinfo{booktitle}{\emph{ICCC}}. \bibinfo{publisher}{Springer},
  \bibinfo{pages}{1--17}.
\newblock


\bibitem[\protect\citeauthoryear{Kossinets and Watts}{Kossinets and
  Watts}{2006}]%
        {kossinets06}
\bibfield{author}{\bibinfo{person}{Gueorgi Kossinets} {and}
  \bibinfo{person}{Duncan~J Watts}.} \bibinfo{year}{2006}\natexlab{}.
\newblock \showarticletitle{Empirical analysis of an evolving social network}.
\newblock \bibinfo{journal}{\emph{Science}} \bibinfo{volume}{311},
  \bibinfo{number}{5757}, \bibinfo{pages}{88--90}.
\newblock


\bibitem[\protect\citeauthoryear{Kumar, Raghavan, Rajagopalan, Sivakumar,
  Tomkins, and Upfal}{Kumar et~al\mbox{.}}{2000}]%
        {kumar00}
\bibfield{author}{\bibinfo{person}{Ravi Kumar}, \bibinfo{person}{Prabhakar
  Raghavan}, \bibinfo{person}{Sridhar Rajagopalan}, \bibinfo{person}{D
  Sivakumar}, \bibinfo{person}{Andrew Tomkins}, {and} \bibinfo{person}{Eli
  Upfal}.} \bibinfo{year}{2000}\natexlab{}.
\newblock \showarticletitle{Stochastic models for the web graph}. In
  \bibinfo{booktitle}{\emph{Proceedings of the 42st Annual Symposium on
  Foundations of Computer Science}}. \bibinfo{publisher}{IEEE},
  \bibinfo{pages}{57--65}.
\newblock


\bibitem[\protect\citeauthoryear{Langholz and Borgan}{Langholz and
  Borgan}{1995}]%
        {langholz95}
\bibfield{author}{\bibinfo{person}{B Langholz} {and} \bibinfo{person}{{\O}rnulf
  Borgan}.} \bibinfo{year}{1995}\natexlab{}.
\newblock \showarticletitle{{Counter-matching: a stratified nested case-control
  sampling method}}.
\newblock \bibinfo{journal}{\emph{Biometrika}} \bibinfo{volume}{82},
  \bibinfo{number}{1}, \bibinfo{pages}{69--79}.
\newblock


\bibitem[\protect\citeauthoryear{Lerner and Lomi}{Lerner and Lomi}{2020}]%
        {lerner19}
\bibfield{author}{\bibinfo{person}{J{\"u}rgen Lerner} {and}
  \bibinfo{person}{Alessandro Lomi}.} \bibinfo{year}{2020}\natexlab{}.
\newblock \showarticletitle{{Reliability of relational event model estimates
  under sampling: how to fit a relational event model to 360 million dyadic
  events}}.
\newblock \bibinfo{journal}{\emph{Network Science}} \bibinfo{volume}{8},
  \bibinfo{number}{1}, \bibinfo{pages}{97--135}.
\newblock


\bibitem[\protect\citeauthoryear{Leskovec, Backstrom, Kumar, and
  Tomkins}{Leskovec et~al\mbox{.}}{2008}]%
        {leskovec08}
\bibfield{author}{\bibinfo{person}{Jure Leskovec}, \bibinfo{person}{Lars
  Backstrom}, \bibinfo{person}{Ravi Kumar}, {and} \bibinfo{person}{Andrew
  Tomkins}.} \bibinfo{year}{2008}\natexlab{}.
\newblock \showarticletitle{{Microscopic evolution of social networks}}. In
  \bibinfo{booktitle}{\emph{KDD}}. \bibinfo{publisher}{ACM},
  \bibinfo{pages}{462--470}.
\newblock


\bibitem[\protect\citeauthoryear{Leskovec and Horvitz}{Leskovec and
  Horvitz}{2008}]%
        {leskovec08b}
\bibfield{author}{\bibinfo{person}{Jure Leskovec} {and} \bibinfo{person}{Eric
  Horvitz}.} \bibinfo{year}{2008}\natexlab{}.
\newblock \showarticletitle{Planetary-scale views on a large instant-messaging
  network}. In \bibinfo{booktitle}{\emph{WWW}}. \bibinfo{publisher}{ACM},
  \bibinfo{pages}{915--924}.
\newblock


\bibitem[\protect\citeauthoryear{Liben-Nowell and Kleinberg}{Liben-Nowell and
  Kleinberg}{2007}]%
        {liben07}
\bibfield{author}{\bibinfo{person}{David Liben-Nowell} {and}
  \bibinfo{person}{Jon Kleinberg}.} \bibinfo{year}{2007}\natexlab{}.
\newblock \showarticletitle{The link-prediction problem for social networks}.
\newblock \bibinfo{journal}{\emph{Journal of the American society for
  information science and technology}} \bibinfo{volume}{58},
  \bibinfo{number}{7}, \bibinfo{pages}{1019--1031}.
\newblock


\bibitem[\protect\citeauthoryear{Luce}{Luce}{1959}]%
        {luce59}
\bibfield{author}{\bibinfo{person}{R~Duncan Luce}.}
  \bibinfo{year}{1959}\natexlab{}.
\newblock \bibinfo{booktitle}{\emph{Individual Choice Behavior; a Theoretical
  Analysis}}.
\newblock \bibinfo{publisher}{New York, Wiley}.
\newblock


\bibitem[\protect\citeauthoryear{Manski and McFadden}{Manski and
  McFadden}{1981}]%
        {manski81}
\bibfield{author}{\bibinfo{person}{C~F Manski} {and} \bibinfo{person}{Daniel
  McFadden}.} \bibinfo{year}{1981}\natexlab{}.
\newblock \showarticletitle{{Alternative estimators and sample designs for
  discrete choice analysis}}.
\newblock In \bibinfo{booktitle}{\emph{Structural Analysis of Discrete Data and
  Econometric Applications}}. \bibinfo{pages}{2--50}.
\newblock


\bibitem[\protect\citeauthoryear{McFadden}{McFadden}{1977}]%
        {mcfadden77}
\bibfield{author}{\bibinfo{person}{Daniel McFadden}.}
  \bibinfo{year}{1977}\natexlab{}.
\newblock \bibinfo{booktitle}{\emph{Modelling the Choice of Residential
  Location}}.
\newblock \bibinfo{type}{Cowles Foundation Discussion Papers} 477.
  \bibinfo{institution}{Yale University}.
\newblock


\bibitem[\protect\citeauthoryear{McPherson, Smith-Lovin, and Cook}{McPherson
  et~al\mbox{.}}{2001}]%
        {mcpherson01}
\bibfield{author}{\bibinfo{person}{Miller McPherson}, \bibinfo{person}{Lynn
  Smith-Lovin}, {and} \bibinfo{person}{James~M Cook}.}
  \bibinfo{year}{2001}\natexlab{}.
\newblock \showarticletitle{{Birds of a feather: Homophily in social
  networks}}.
\newblock \bibinfo{journal}{\emph{Annual Review of Sociology}}
  \bibinfo{volume}{32}, \bibinfo{number}{1}, \bibinfo{pages}{19--28}.
\newblock


\bibitem[\protect\citeauthoryear{Mikolov, Sutskever, Chen, Corrado, and
  Dean}{Mikolov et~al\mbox{.}}{2013}]%
        {mikolov13}
\bibfield{author}{\bibinfo{person}{Tomas Mikolov}, \bibinfo{person}{Ilya
  Sutskever}, \bibinfo{person}{Kai Chen}, \bibinfo{person}{Greg~S Corrado},
  {and} \bibinfo{person}{Jeff Dean}.} \bibinfo{year}{2013}\natexlab{}.
\newblock \showarticletitle{Distributed representations of words and phrases
  and their compositionality}. In \bibinfo{booktitle}{\emph{NeurIPS}}.
  \bibinfo{publisher}{Curran Associates}, \bibinfo{pages}{3111--3119}.
\newblock


\bibitem[\protect\citeauthoryear{Nerella and Bhat}{Nerella and Bhat}{2004}]%
        {nerella04}
\bibfield{author}{\bibinfo{person}{S Nerella} {and} \bibinfo{person}{C Bhat}.}
  \bibinfo{year}{2004}\natexlab{}.
\newblock \showarticletitle{{Numerical analysis of effect of sampling of
  alternatives in discrete choice models}}.
\newblock \bibinfo{journal}{\emph{J of Transportation Research}}
  \bibinfo{volume}{1894}, \bibinfo{number}{1}, \bibinfo{pages}{11--19}.
\newblock


\bibitem[\protect\citeauthoryear{Nesterov}{Nesterov}{2013}]%
        {nesterov13}
\bibfield{author}{\bibinfo{person}{Yurii Nesterov}.}
  \bibinfo{year}{2013}\natexlab{}.
\newblock \bibinfo{booktitle}{\emph{Introductory lectures on convex
  optimization: A basic course}}. Vol.~\bibinfo{volume}{87}.
\newblock \bibinfo{publisher}{Springer Science \& Business Media}.
\newblock


\bibitem[\protect\citeauthoryear{Noble and Cook}{Noble and Cook}{2003}]%
        {noble03}
\bibfield{author}{\bibinfo{person}{Caleb~C Noble} {and}
  \bibinfo{person}{Diane~J Cook}.} \bibinfo{year}{2003}\natexlab{}.
\newblock \showarticletitle{Graph-based anomaly detection}. In
  \bibinfo{booktitle}{\emph{KDD}}. \bibinfo{publisher}{ACM},
  \bibinfo{pages}{631--636}.
\newblock


\bibitem[\protect\citeauthoryear{Overgoor, Benson, and Ugander}{Overgoor
  et~al\mbox{.}}{2019}]%
        {overgoor19}
\bibfield{author}{\bibinfo{person}{Jan Overgoor}, \bibinfo{person}{Austin
  Benson}, {and} \bibinfo{person}{Johan Ugander}.}
  \bibinfo{year}{2019}\natexlab{}.
\newblock \showarticletitle{Choosing to Grow a Graph: Modeling Network
  Formation As Discrete Choice}. In \bibinfo{booktitle}{\emph{WWW}}.
  \bibinfo{publisher}{ACM}, \bibinfo{pages}{1409--1420}.
\newblock


\bibitem[\protect\citeauthoryear{Owen}{Owen}{2013}]%
        {owen13}
\bibfield{author}{\bibinfo{person}{Art~B. Owen}.}
  \bibinfo{year}{2013}\natexlab{}.
\newblock \bibinfo{booktitle}{\emph{Monte Carlo theory, methods and examples}}.
\newblock


\bibitem[\protect\citeauthoryear{Rapoport}{Rapoport}{1953}]%
        {rapoport53}
\bibfield{author}{\bibinfo{person}{Anatol Rapoport}.}
  \bibinfo{year}{1953}\natexlab{}.
\newblock \showarticletitle{Spread of information through a population with
  socio-structural bias: I. Assumption of transitivity}.
\newblock \bibinfo{journal}{\emph{The bulletin of mathematical biophysics}}
  \bibinfo{volume}{15}, \bibinfo{number}{4}, \bibinfo{pages}{523--533}.
\newblock


\bibitem[\protect\citeauthoryear{Robins, Pattison, Kalish, and Lusher}{Robins
  et~al\mbox{.}}{2007}]%
        {robins07}
\bibfield{author}{\bibinfo{person}{Garry Robins}, \bibinfo{person}{Pip
  Pattison}, \bibinfo{person}{Yuval Kalish}, {and} \bibinfo{person}{Dean
  Lusher}.} \bibinfo{year}{2007}\natexlab{}.
\newblock \showarticletitle{An introduction to exponential random graph (p*)
  models for social networks}.
\newblock \bibinfo{journal}{\emph{Social networks}} \bibinfo{volume}{29},
  \bibinfo{number}{2}, \bibinfo{pages}{173--191}.
\newblock


\bibitem[\protect\citeauthoryear{Seshadri, Peysakhovich, and Ugander}{Seshadri
  et~al\mbox{.}}{2019}]%
        {seshadri19}
\bibfield{author}{\bibinfo{person}{Arjun Seshadri}, \bibinfo{person}{Alex
  Peysakhovich}, {and} \bibinfo{person}{Johan Ugander}.}
  \bibinfo{year}{2019}\natexlab{}.
\newblock \showarticletitle{Discovering Context Effects from Raw Choice Data}.
  In \bibinfo{booktitle}{\emph{ICML}}. \bibinfo{publisher}{PMLR},
  \bibinfo{pages}{5660--5669}.
\newblock


\bibitem[\protect\citeauthoryear{Snijders, Van~de Bunt, and Steglich}{Snijders
  et~al\mbox{.}}{2010}]%
        {snijders10}
\bibfield{author}{\bibinfo{person}{Tom~AB Snijders}, \bibinfo{person}{Gerhard~G
  Van~de Bunt}, {and} \bibinfo{person}{Christian~EG Steglich}.}
  \bibinfo{year}{2010}\natexlab{}.
\newblock \showarticletitle{Introduction to stochastic actor-based models for
  network dynamics}.
\newblock \bibinfo{journal}{\emph{Social networks}} \bibinfo{volume}{32},
  \bibinfo{number}{1}, \bibinfo{pages}{44--60}.
\newblock


\bibitem[\protect\citeauthoryear{Stadtfeld and Block}{Stadtfeld and
  Block}{2017}]%
        {stadtfeld17a}
\bibfield{author}{\bibinfo{person}{Christoph Stadtfeld} {and}
  \bibinfo{person}{Per Block}.} \bibinfo{year}{2017}\natexlab{}.
\newblock \showarticletitle{{Interactions, Actors, and Time: Dynamic Network
  Actor Models for Relational Events}}.
\newblock \bibinfo{journal}{\emph{Sociological Science}}  \bibinfo{volume}{4},
  \bibinfo{pages}{318--352}.
\newblock


\bibitem[\protect\citeauthoryear{Tang, Qu, Wang, Zhang, Yan, and Mei}{Tang
  et~al\mbox{.}}{2015}]%
        {tang15}
\bibfield{author}{\bibinfo{person}{Jian Tang}, \bibinfo{person}{Meng Qu},
  \bibinfo{person}{Mingzhe Wang}, \bibinfo{person}{Ming Zhang},
  \bibinfo{person}{Jun Yan}, {and} \bibinfo{person}{Qiaozhu Mei}.}
  \bibinfo{year}{2015}\natexlab{}.
\newblock \showarticletitle{Line: Large-scale information network embedding}.
  In \bibinfo{booktitle}{\emph{WWW}}. \bibinfo{publisher}{ACM},
  \bibinfo{pages}{1067--1077}.
\newblock


\bibitem[\protect\citeauthoryear{Thurstone}{Thurstone}{1927}]%
        {thurstone27}
\bibfield{author}{\bibinfo{person}{Louis~L Thurstone}.}
  \bibinfo{year}{1927}\natexlab{}.
\newblock \showarticletitle{A law of comparative judgment}.
\newblock \bibinfo{journal}{\emph{Psychological review}} \bibinfo{volume}{34},
  \bibinfo{number}{4}, \bibinfo{pages}{273--286}.
\newblock


\bibitem[\protect\citeauthoryear{Train}{Train}{2009}]%
        {train09}
\bibfield{author}{\bibinfo{person}{Kenneth~E Train}.}
  \bibinfo{year}{2009}\natexlab{}.
\newblock \bibinfo{booktitle}{\emph{Discrete choice methods with simulation}}.
\newblock \bibinfo{publisher}{CUP}.
\newblock


\bibitem[\protect\citeauthoryear{Tukey}{Tukey}{1960}]%
        {tukey60}
\bibfield{author}{\bibinfo{person}{John~W Tukey}.}
  \bibinfo{year}{1960}\natexlab{}.
\newblock \showarticletitle{A survey of sampling from contaminated
  distributions}.
\newblock \bibinfo{journal}{\emph{Contributions to probability and
  statistics}}, \bibinfo{pages}{448--485}.
\newblock


\bibitem[\protect\citeauthoryear{Ugander and Backstrom}{Ugander and
  Backstrom}{2013}]%
        {ugander13}
\bibfield{author}{\bibinfo{person}{Johan Ugander} {and} \bibinfo{person}{Lars
  Backstrom}.} \bibinfo{year}{2013}\natexlab{}.
\newblock \showarticletitle{Balanced label propagation for partitioning massive
  graphs}. In \bibinfo{booktitle}{\emph{WSDM}}. \bibinfo{publisher}{ACM},
  \bibinfo{pages}{507--516}.
\newblock


\bibitem[\protect\citeauthoryear{von Haefen and Domanski}{von Haefen and
  Domanski}{2018}]%
        {vonhaefen18}
\bibfield{author}{\bibinfo{person}{Roger~H von Haefen} {and}
  \bibinfo{person}{Adam Domanski}.} \bibinfo{year}{2018}\natexlab{}.
\newblock \showarticletitle{{Estimation and welfare analysis from mixed logit
  models with large choice sets}}.
\newblock \bibinfo{journal}{\emph{Journal of Environmental Economics and
  Management}}  \bibinfo{volume}{90}, \bibinfo{pages}{101--118}.
\newblock


\bibitem[\protect\citeauthoryear{Vu, Pattison, and Robins}{Vu
  et~al\mbox{.}}{2015}]%
        {vu15}
\bibfield{author}{\bibinfo{person}{Duy Vu}, \bibinfo{person}{Philippa
  Pattison}, {and} \bibinfo{person}{Garry Robins}.}
  \bibinfo{year}{2015}\natexlab{}.
\newblock \showarticletitle{{Relational event models for social learning in
  MOOCs}}.
\newblock \bibinfo{journal}{\emph{Social Networks}}  \bibinfo{volume}{43},
  \bibinfo{pages}{121--135}.
\newblock


\bibitem[\protect\citeauthoryear{Watts}{Watts}{2003}]%
        {watts03}
\bibfield{author}{\bibinfo{person}{Duncan~J Watts}.}
  \bibinfo{year}{2003}\natexlab{}.
\newblock \bibinfo{booktitle}{\emph{Small worlds: the dynamics of networks
  between order and randomness}}.
\newblock \bibinfo{publisher}{Princeton university press}.
\newblock


\bibitem[\protect\citeauthoryear{Zhang, Tang, Zhao, Wang, Zheng, and
  Zhao}{Zhang et~al\mbox{.}}{2017}]%
        {zhang17}
\bibfield{author}{\bibinfo{person}{Xinyi Zhang}, \bibinfo{person}{Shiliang
  Tang}, \bibinfo{person}{Yun Zhao}, \bibinfo{person}{Gang Wang},
  \bibinfo{person}{Haitao Zheng}, {and} \bibinfo{person}{Ben~Y Zhao}.}
  \bibinfo{year}{2017}\natexlab{}.
\newblock \showarticletitle{Cold hard E-cash: Friends and vendors in the Venmo
  digital payments system}. In \bibinfo{booktitle}{\emph{ICWSM}}.
  \bibinfo{publisher}{ACM}, \bibinfo{pages}{387--396}.
\newblock


\end{thebibliography}

\end{document}